\documentclass[submission]{eptcs}
\providecommand{\event}{FMDS 2012} 
\usepackage{breakurl}             

\usepackage{stmaryrd}
\usepackage{amssymb}
\usepackage{amsmath}
\usepackage{enumerate}
\usepackage[algoruled,vlined,linesnumbered]{algorithm2e}

\usepackage{tikz}
\usetikzlibrary{arrows,automata,shapes}

\usepackage{ntheorem}
\theoremstyle{definition}
\newtheorem{definition}{Definition}
\newtheorem{example}{Example}
\theoremstyle{plain}
\newtheorem{theorem}{Theorem}

\newcounter{soscounter}

\newcommand{\pp}{\setdisp{R}}

\newcommand{\s}{\fndisp{S_B}}

\newcounter{axcounter}

\newcommand{\bbbn}{\mathbb{N}}

\newcommand{\setdisp}[1]{\mathsf{#1}}
\newcommand{\fndisp}[1]{\mathrm{#1}}

\newcommand{\pcal}{\setdisp{P}}
\newcommand{\ocal}{\setdisp{O}}

\newcommand{\Sact}{\setdisp{A}}
\newcommand{\Slab}{\setdisp{A}}
\newcommand{\Sst}{\setdisp{S}}

\newcommand{\Cterms}{\setdisp{S}}

\newcommand{\pb}[1]{\mathbin{\leq}}

\newcommand{\bpb}[1]{\mathbin{=}}

\newcommand{\pbis}[1]{\mathbin{\leq_{#1}}}

\newcommand{\bpbis}[1]{\mathbin{\leftrightarrow_{#1}}}
\newcommand{\vbpbis}[1]{\mathbin{\leftrightarrow_{#1}}}
\newcommand{\lb}{\mathbin{\sqsubseteq}}

\newcommand{\bbc}[2]{#1_{#2}}
\newcommand{\lbc}[2]{{}_{#2}\!#1}

\newcommand{\pcl}[1]{[#1]_{\bpbis{B}}}
\newcommand{\vpcl}[1]{[#1]_{\bpbis{B}}}

\newcommand{\tss}[1]{#1\mkern1mu \mathord{\downarrow}}
\newcommand{\nts}[1]{#1\mkern1mu \mathord{\not \mkern2mu\downarrow}}
\newcommand{\tr}[1]{\mathbin{\stackrel{#1}{\mathord{\rightarrow}}}}

\newcommand{\aetr}[1]{\mathbin{\stackrel{#1}{\mathord{\rightarrow}}_{\exists\forall}}}

\newcommand{\extr}[1]{\mathbin{\stackrel{#1}{\mathord{\rightarrow}}_\exists}}
\newcommand{\altr}[1]{\mathbin{\stackrel{#1}{\mathord{\rightarrow}}_\forall}}

\newcommand{\cntlb}{\fndisp{cnt}_{\lb}}
\newcommand{\cntal}{\fndisp{cnt}_{\forall}}

\title{Scalable Minimization Algorithm for Partial Bisimulation}
\author{J. Markovski
\institute{Eindhoven University of Technology, \\PB 513, 5600MB, Eindhoven, The Netherlands, }
\email{j.markovski@tue.nl}
}

\def\titlerunning{Scalable Minimization Algorithm for Partial Bisimulation}
\def\authorrunning{J. Markovski}
\begin{document}
\maketitle

\begin{abstract}
We present an efficient algorithm for computing the partial bisimulation preorder and equivalence for labeled transitions systems. The partial bisimulation preorder lies between simulation and bisimulation, as only a part of the set of actions is bisimulated, whereas the rest of the actions are simulated. Computing quotients for simulation equivalence is more expensive than for bisimulation equivalence, as for simulation one has to account for the so-called little brothers, which represent classes of states that can simulate other classes. It is known that in the absence of little brother states, (partial bi)simulation and bisimulation coincide, but still the complexity of existing minimization algorithms for simulation and bisimulation does not scale. Therefore, we developed a minimization algorithm and an accompanying tool that scales with respect to the bisimulated action subset.
\end{abstract}

\section{Introduction}

A recent process-theoretic approach to supervisory control theory~\cite{acc2011} identified partial bisimulation preorder~\cite{coalgebra} as a suitable behavioral equivalence that captures the central notion of controllability~\cite{rwsupervisor,Cassandras}. The property of controllability conditions automated synthesis of supervisory control software based on the discrete-event models of the uncontrolled system and the control requirements. Supervisory controllers ensure safe and nonblocking behavior of the supervised system with respect to a given set of control requirements. Safe behavior means that the supervised system remains in the domain of the control requirements, whereas nonblocking behavior is achieved by removing deadlock or livelock~\cite{rwsupervisor,Cassandras}.

The supervisory controllers observe the discrete behavior of the uncontrolled system by receiving sensor signals, make a decision on which activities the system can safely perform, and send back control signals that actuate the system. Typically, it is assumed that the supervisory controller can react sufficiently fast on machine input, which enables modeling of the supervisory control loop as a pair of synchronizing processes. Then, controllability condition states that the supervisory controller must never disable sensor events, also known as uncontrollable events, in order to achieve the control requirements. Instead, it can only disable actuator signals, known as controllable events, so that the behavior of the system remains safely within the bounds of the control requirements and it is nonblocking.

In process-theoretic terms, the model of the uncontrolled system can be viewed as a specification, whereas the model of the supervised system is an implementation. Then, partial bisimulation preorder is established in such a way that the specification simulates the controllable events of the implementation, whereas all reachable states with outgoing uncontrollable events must be bisimulated in order to ensure that the supervisory controller does not disabled them. During the synthesis process, the control requirements often change as designers develop the product, whereas the model of the uncontrolled system, i.e., the hardware, remains fixed. Therefore, it is of interest to minimize the model of the uncontrolled system with respect to induced partial bisimulation equivalence in order to optimize the synthesis procedure.

The partial bisimulation equivalence is parameterized with a bisimulation action set that identifies the labels of the transitions that are to be bisimulated. If the bisimulation action set is empty, then partial bisimulation equivalence reduces to simulation equivalence~\cite{glabbeek,acc2011}. If the bisimulation set comprises all action labels, then partial bisimulation equivalence reduces to bisimulation equivalence~\cite{glabbeek,acc2011}. For any other bisimulation action set, the corresponding partial bisimulation equivalence lies between these two equivalences. There exist efficient proposals for minimization algorithms for both simulation~\cite{partitionpair,bustan,henzingersimulation,ploeger,efficient-ranzato,FMDS2011} and bisimulation equivalences~\cite{paigetarjan,fernandez,katoen-modelchecking}. Suppose that the system to be minimized has a set of states~$\Sst$, a transition relation~$\tr{}$, a set of action labels~$\Sact$, and resulting partition classes contained in partition~$\pcal$. Then, the most efficient minimization algorithm for simulation has time complexity of $\ocal(|\pcal||\mathord{\tr{}}|)$~\cite{henzingersimulation}, whereas most efficient bisimulation algorithm for bisimulation has time complexity $\ocal(|\mathord{\tr{}}|\log(|\Sst|))$~\cite{paigetarjan,fernandez}. Moreover, the minimization algorithm for simulation that offers the best compromise between time and space complexity has time complexity of $\ocal(|\pcal||\Sst|\log(|\Sst|))$ and space complexity of $\ocal(|\Sst|\log(|\pcal|) + |\pcal|^2)$~\cite{efficient-ranzato,FMDS2011}.

The discrepancy between the minimization algorithms for bisimulation and simulation lies in the fact that for simulation, one has to additionally account for the so-called little brother relation, which relates partition classes that can simulate each other~\cite{partitionpair}. It has been shown that if the little brother relation is empty, then simulation and bisimulation actually coincide~\cite{katoen-modelchecking,report-concur}. However, if we observe the complexities of the minimization algorithms, we can easily observe that the algorithms do not scale accordingly. This is the result of computing the state space partition and the little brother relation simultaneously in order to increase overall time efficiency~\cite{efficient-ranzato}. When dealing solely with simulation, this proves a valid strategy. However, for partial bisimulation, we propose to decouple the computation of the underlying partition from the updating the little brother relation, thus obtaining a scalable implementation. For the former, we employ techniques from bisimulation minimization~\cite{paigetarjan,fernandez}, while for the latter we rely on the representation of the little brother relation of~\cite{FMDS2011}. The resulting minimization algorithm has worst-case time complexity of $\ocal(|\mathord{\tr{}}|\log(|\Sst|) + |\Sact||\pcal||\mathord{\lb}|)$, where $\lb$ is the little brother relation, while having a space complexity of $\ocal(|\Sact||\Sst|\log(|\pcal|) + |\Sact||\pcal|^2\log(|\pcal|)$. We note a slight increase of space complexity with respect to~\cite{efficient-ranzato,FMDS2011} as we employ an additional set of counters that optimize the partition splitting in the vein of~\cite{paigetarjan,fernandez}.

The rest of this paper is organized as follows. In section 2, we revisit the notion of partial bisimulation and discuss an alternative representation in the form of a partition-relation pair. Afterwards, in section 3 we develop a refinement for partition-relation pairs that results in the coarsest partial bisimulation quotient. We discuss the implementation of the algorithm in section 4 and finish with concluding remarks. We note that a prototype implementation of the algorithm can be downloaded from~\cite{website}, whereas technical details and proofs are given in~\cite{report-concur}.

\section{Partial Bisimulation and Partition-Relation Pairs Representation}

The underlying model that we consider is labeled transitions systems (with successful termination options) following the notation of~\cite{pabook,acc2011}. A labeled transition system $G$ is a tuple $G \triangleq (\Sst, \Sact, \tss{}, \tr{})$, where $\Sst$ is a set of states, $\Sact$ a set of event labels, $\tss{} \subseteq \Sst$ is a successful termination predicate that takes the role of marked or final states in supervisory control setting~\cite{rwsupervisor,Cassandras,acc2011}, and $\tr{} \subseteq \Sst \times \Sact \times \Sst$ is the labeled transition relation. For $p, q \in \Sst$ and $a \in \Sact$, we write $p \tr{a} q$ and $\tss{p}$.

\begin{definition}\label{def:partial-bisimilarity}
A relation~$R \subseteq \Sst \times \Sst$ is a partial bisimulation with respect to the bisimulation action set~$B \subseteq \Sact$, if for all $(p,q) \in R$ it holds that:
\begin{enumerate}
\item  if $\tss{p}$, then $\tss{q}$;

\item 
if $p \tr{a} p'$ for some $a \in \Sact$, then there exists $q' \in \Cterms$ such that $q \tr{a} q'$ and $(p',q') \in R$;

\item 
if $q \tr{b} q'$ for some $b \in B$, then there exists $p' \in \Cterms$ such that $p \tr{b} p'$ and $(p',q') \in R$.
\end{enumerate}
If $(p, q) \in R$, then $p$ is partially bisimilar to $q$, notation $p \pbis{B} q$. If $q \pbis{B} p$ holds as well, we write $p \bpbis{B} q$.
\end{definition}
It is not difficult to show that $\pbis{B}$ is a preorder relation, making $\bpbis{B}$ an equivalence relation for all $B\subseteq \Sact$~\cite{acc2011}. If $B=\emptyset$, then $\pbis{\emptyset}$ coincides with strong similarity preorder and $\bpbis{\emptyset}$ coincides with strong similarity equivalence~\cite{glabbeek,pabook}. When $B = \Sact$, $\bpbis{\Sact}$ turns into strong bisimilarity~\cite{glabbeek,pabook}.

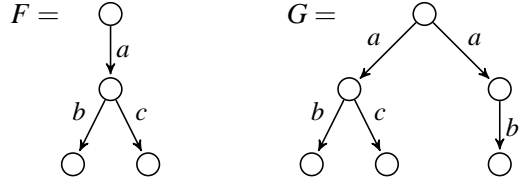
\begin{figure}[t!]
\centering
\begin{tabular}{ccc}
\begin{tikzpicture}[->,>=stealth',shorten >=1pt,%
auto, node distance=1cm, semithick,
inner sep=2pt,bend angle=15, initial text=, initial where = left,
every state/.style={circle, semithick, minimum size = 3mm}]

\node[state] (A) {};
\node (X) [left of=A, node distance = 1cm] {$F = $};
\node[state] (B) [below of=A] {};
\node (C) [below of=B] {};
\node[state] (D) [left of=C, node distance = 0.5cm] {};
\node[state] (E) [right of=C, node distance = 0.5cm] {};

\tikzstyle{every node}=[font = \small]

\path

(A) edge node {\emph{a}} (B)
(B) edge node[swap] {\emph{b}} (D)
(B) edge node {\emph{c}} (E)

%
%
%
%
%
%
%
;
\end{tikzpicture}
& \qquad \qquad &
\begin{tikzpicture}[->,>=stealth',shorten >=1pt,%
auto, node distance=1cm, semithick,
inner sep=2pt,bend angle=15, initial text=, initial where = left,
every state/.style={circle, semithick, minimum size = 3mm}]

\node[state] (A) {};
\node (X) [left of=A, node distance = 1.5cm] {$G = $};
\node (B) [below of=A] {};
\node[state] (BB) [left of=B] {};
\node[state] (BBB) [right of=B] {};
\node[state] (CC) [below of=BBB] {};
\node (C) [below of=BB] {};
\node[state] (D) [left of=C, node distance = 0.5cm] {};
\node[state] (E) [right of=C, node distance = 0.5cm] {};

\tikzstyle{every node}=[font = \small]

\path

(A) edge node[swap] {\emph{a}} (BB)
(A) edge node {\emph{a}} (BBB)
(BBB) edge node {\emph{b}} (CC)
(BB) edge node[swap] {\emph{b}} (D)
(BB) edge node {\emph{c}} (E)
;
\end{tikzpicture}
\end{tabular}
\caption{Partial bisimulation example
} \label{fig:pb-example}
\end{figure}

\begin{example}\label{ex::pb}
To provide better intuition, we consider the labeled transition systems depicted in Figure~\ref{fig:pb-example}. Following Definition~\ref{def:partial-bisimilarity}, if $B = \{a, b, c\}$, then we are looking to establish bisimulation between the labeled transition systems $F$ and $G$, which is not possible. If $B \subseteq \{b\}$, then we have $F \vbpbis{B} G$. For $B = \{c\}$, we cannot establish that $F \pbis{B} G$ as the right branch of $G$ has no outgoing transition labeled by~$c$, whereas it is required that this transition is (bi)simulated back in the partial bisimulation relation.
\end{example}
An important role in simulation-like relations is played by the so-called little brother state pairs~\cite{partitionpair,efficient-ranzato}. Little brother state pairs identify states reachable in an equivalent manner, where one state has strictly greater behavior with respect to the underlying relation. We say that $p'$ is the little brother of $p''$ if $p \tr{a} p'$ and $p \tr{a} p''$ with $p'\pbis{B} p''$. For example, the states reachable by the transition labeled by $a$ in the labeled transition system $G$ in Figure~\ref{fig:pb-example} represent a little brother pair, provided that $c \not \in B$.

If there are no little brother pairs in a simulation relation, then the related processes are actually bisimilar~\cite{katoen-modelchecking,acc2011}. The greatest challenge in minimization procedures for simulation-based relations lies in efficient treatment of the little brother pairs~\cite{partitionpair,efficient-ranzato,katoen-modelchecking,FMDS2011}. The following theorem of~\cite{acc2011,report-concur} shows how to address little brother pairs for partial bisimilarity.

\begin{theorem}\label{thm:little}
Let $p_1 \pbis{B} p_2 \pbis{B} p_3$ for $p_1\mathord{,} p_2\mathord{,} p_3 \in \Sst$, and let $a \in \Sact \setminus B$ and $b \in B$ for $B \subseteq \Sact$.
\begin{enumerate}
\item If $q_1, q_2 \in \Sst$ are such that $q_1 \tr{a} p_1$, $q_1 \tr{a} p_2$, and $q_2 \tr{a} p_2$, 
    then $q_1 \vbpbis{B} q_2$.
\item If $q_1, q_2 \in \Sst$ are such that $q_1 \tr{b} p_1$, $q_1 \tr{b} p_2$, $q_1 \tr{b} p_3$, $q_2 \tr{b} p_1$, and $q_2 \tr{b} p_3$, 
    then $q_1 \vbpbis{B} q_2$.
\end{enumerate}
\end{theorem}
Intuitively, Theorem~\ref{thm:little} states that for controllable events, retaining the biggest brother is sufficient, whereas for uncontrollable events, both the littlest and the biggest brother must be preserved.

To optimize the computation of the little brother pairs, in the sequel we represent partial bisimilarity preorders by means of partition-relation pairs~\cite{partitionpair}. The partition identifies equivalent states by partial bisimilarity, which are placed into the same classes, whereas the relation, given between the partition classes, identifies the little brother pairs in the partition, thus forming the quotient. Let $G = (\Sst, \Slab, \tss{}, \tr{})$ and let $\pcal \subset 2^\Sst$. The set $\pcal$ is a partition over $\Sst$ if $\bigcup_{P \in \pcal} P = \Sst$ and for all $P, Q \in \pcal$, if $P \cap Q \neq \emptyset$, then $P = Q$. A partition-relation pair over $G$ is a pair $(\pcal, \lb)$ where $\pcal$ is a partition over~$\Sst$ and the (little brother) relation $\lb \subseteq \pcal \times \pcal$ is a partial order, i.e., a reflexive, antisymmetric, and transitive relation. We denote the set of partition-relation pairs by $\pp$.

The partition classes induce several (Galois) relations relying on $\downarrow$ and $\tr{}$~\cite{partitionpair}. For all $P \in \pcal$, we have that $\tss{P}$ or $\nts{P}$, if for all $p \in P$ it holds that $\tss{p}$ or $\nts{p}$, respectively. For all $P' \in \pcal$ by $p \tr{a} P'$ we denote that there exists $p' \in P'$ such that $p \tr{a} p'$. Moreover, by $P \extr{a} P'$ we denote that there exists $p \in P$ such that $p \tr{a} P'$, and by $P \altr{a} P'$ we denote that for every $p \in P$, it holds that $p \tr{a} P'$. It is straightforward that $P \altr{a} P'$ implies $P \extr{a} P'$. Also, if $P \altr{a} P'$, then $Q \altr{a} P'$ for every $Q \subseteq P$.

To relate partial bisimulation preorders and partition-relation pairs, we rely on stability conditions which must hold for a given pair, so that it induces a partial bisimilarity preorder with respect to the termination predicate and the transition relation. Vice versa, we show that every partial bisimulation preorder induces a stable partition-relation pair. To this end, we define by $\lbc{P}{\lb}\triangleq \bigcup \{Q \in \pcal \mid Q \lb P\}$ and $\bbc{P}{\lb} \triangleq \bigcup \{Q \in \pcal \mid P \lb Q\}$ all little and big brother classes of the partition class $P \in \pcal$, respectively. Also, given a relation~$R \in S \times T$ on some sets $S$ and $T$, we define $R^{-1} \in T \times S$ as $R^{-1} \triangleq \{ (t,s) \mid (s,t) \in R\}$. Moreover, we note that if a given relation $R$ is a preorder, then $R \cap R^{-1}$ is an equivalence relation. If $\bpbis{}$ is an equivalence over $\Sst$, then $\Sst/\bpbis{}$ denotes the induced partition, whereas $[p]_{\bpbis{}}$ is the partition class of $p \in \Sst$. First, we define the stability conditions that ensure that a partition-relation pair induces a partial bisimulation preorder.

\begin{definition}\label{def:stabilityconditions}
Let~$G = (\Sst, \Slab, \tss{}, \tr{})$ be a labeled graph. We say that $(\pcal,\lb) \in \pp$ over~$G$ is stable (with respect to $\tss{}$, $\tr{}$, and $B\subseteq \Sact$) if the following conditions are fulfilled:
\begin{itemize}
\item[a.] For all $P \in \pcal$, it holds that $\tss{P}$ or $\nts{P}$.

\item[b.] For all $P, Q \in \pcal$, if $P \lb Q$ and $\tss{P}$, then $\tss{Q}$.

\item[c.] For all $P, Q, R \in \pcal$ and $a \in \Sact$, if $P \lb Q$ and $P \extr{a} R$, then $Q \altr{a} \bbc{R}{\lb}$.

\item[d.] For all $P, Q, R \in \pcal$ and $b \in B$, if $P \lb Q$ and $Q \extr{b} R$, then $P \altr{b} \lbc{R}{\lb}$.

\end{itemize}
\end{definition}
Having in mind Definition~\ref{def:partial-bisimilarity}, conditions \emph{a} and \emph{b} require that partially bisimilar equivalent states must have the same termination options, whereas big brothers must be able to terminate if the little brother is able to terminate. Condition \emph{c} corresponds to the stability condition for simulation~\cite{partitionpair,efficient-ranzato,FMDS2011} and it states that if a little brother can perform a transition labeled by $a \in \Sact$, then the big brother must also enable such a transition, possibly ending in a big brother of the target class. Condition \emph{d} is actually induced by Theorem~\ref{thm:little} and it states that every little brother must be able to follow transitions labeled by $b \in B$ that are enabled by a big brother, possibly ending in a little brother of the target class.

Next, we show that every partial bisimulation preorder induced a stable partition-relation pair~\cite{report-concur}.

\begin{theorem}\label{thm:pb2stability}
Let $G = (\Sst, \Sact, \tss{}, \tr{})$ and let~$R$ be a partial bisimulation preorder over~$\Sst$ with respect to $B \subseteq \Sact$. Let $\bpbis{B} \triangleq R \cap R^{-1}$. If $\pcal = \Sst / \bpbis{B}$ and $\lb \subseteq \pcal \times \pcal$ are such that for all $(p, q) \in R$ it holds $\pcl{p} \lb \pcl{q}$, then $(\pcal,\lb) \in \pp$ is stable.
\end{theorem}
Vice versa, stable partition-relation pairs induce partial bisimulation preorders.

\begin{theorem}\label{thm:stability2pb}
Let $G = (\Sst, \Sact, \tss{}, \tr{})$ and $(\pcal,\lb) \in \pp$. Define $R = \{ (p,q) \in P \times Q \mid P \lb Q\}$. If $(\pcal,\lb)$ is stable, then $R$ is a partial bisimulation preorder for $B$.
\end{theorem}
Theorems~\ref{thm:pb2stability} and \ref{thm:stability2pb} enable us to refine partition-relation pairs instead of dealing directly with the partial bisimulation preorder. We specify a fix-point refinement operator that induces the coarsest stable partition-relation pair that induces the greatest partial bisimulation preorder and equivalence.

\section{Refinement Operator}

To define the refinement operator, we need to specify when we consider one partition-relation pair to be finer than another pair. Moreover, finer stable partition-relation pairs should correspond to finer induced partial bisimulation preorders.

\begin{definition}\label{def:partialorderpartitions}
Let $(\pcal,\lb)$ and $(\pcal',\lb')$ be partition-relation pairs. We say that $(\pcal,\lb)$ is finer than $(\pcal',\lb')$, notation $(\pcal,\lb) \lhd (\pcal',\lb')$, if and only if for all $P, Q \in \pcal$ such that $P \lb Q$ there exist $P'\!, Q' \in \pcal'$ such that $P \subseteq P'\!$, $Q \subseteq Q'\!$, and $P' \lb' Q'$.
\end{definition}
The relation $\lhd$ as given in Definition~\ref{def:partialorderpartitions} is a partial order~\cite{report-concur}. The following theorem states that coarser partition-relation pairs with respect to $\lhd$ produce coarser partial bisimulation preorders.

\begin{theorem}\label{thm:lhd-induced-subset}
Let $G = (\Sst, \Sact, \tss{}, \tr{})$ and $(\pcal_1,\lb_1)$, $(\pcal_2,\lb_2)$ $ \in \pp$. Define $R_i = \{ (p_i,q_i) \in P_i \times Q_i \mid P_i \lb_i Q_i\}$ for $i \in \{1,2\}$. Then $(\pcal_1,\lb_1) \lhd (\pcal_2,\lb_2)$ if and only if $R_1 \subseteq R_2$.
\end{theorem}
For the refinement operator to contain a unique fix point, we have to establish a confluence property, i.e., for every two stable partition pairs of the same graph, there exists a $\lhd$-coarser stable partition pair.

\begin{theorem}\label{thm:lhd-lattice}
Let $G = (\Sst, \Slab, \tss{}, \tr{})$ and let $(\pcal_1, \lb_1)$, $(\pcal_2, \lb_2)$ $\in \pp$ be stable partition pairs. Then, there exists $(\pcal_3, \lb_3) \in \pp$ that is also stable, and $(\pcal_1, \lb_1)\lhd (\pcal_3, \lb_3)$ and $(\pcal_2, \lb_2)\lhd (\pcal_3, \lb_3)$.
\end{theorem}
Theorem~\ref{thm:lhd-lattice} implies that stable partition pairs form an upper lattice with respect to $\lhd$. Now, it is not difficult to observe that finding the $\lhd$-maximal stable partition pair over a labeled graph~$G$ coincides with the problem of finding the coarsest partial bisimulation preorder over~$G$.

\begin{theorem}\label{thm:lhd-maximal}
Let $G = (\Sst, \Sact, \tss{}, \tr{})$. The $\lhd$-maximal $(\pcal, \lb) \in \pp$ that is stable is induced by the partial bisimilarity preorder~$\pbis{B}$, i.e., $\pcal = \Sst / \bpbis{B}$ and $\vpcl{p} \lb \vpcl{q}$ if and only if $p \pbis{B} q$.
\end{theorem}
Theorem~\ref{thm:lhd-maximal} supported by Theorem~\ref{thm:lhd-lattice} induces an algorithm for computing the coarsest mutual partial bisimulation over a labeled transition system $G = (\Sst, \Sact, \tss{}, \tr{})$ by computing the $\lhd$-maximal partition pair $(\pcal, \lb)$ such that $(\pcal, \lb) \lhd (\{\Sst\}, \{(\Sst, \Sst)\})$. We develop an iterative algorithm that refines this partition pair, until it reaches the $\lhd$-maximal stable partition pair.

The algorithm works in two phases. First, we refine the partition, followed by an update of the partition pair. We refine the partitions by splitting them in the vein of~\cite{paigetarjan,fernandez,katoen-modelchecking}, i.e., we choose subsets of states, referred to as splitters, that do not adhere to the stability conditions in combination with the other states from the same class and, consequently, we place them in a separate class. To this end we distinguish between parent partitions and child partitions, the former comprising the potential splitters, whereas the latter hold the result of the current application of the refinement algorithm.

Let $(\pcal, \lb) \in \pp$ be defined over $\Sst$. Partition $\pcal'$ is a parent partition of $\pcal$, if for every $P \in \pcal$, there exist $P' \in \pcal'$ with $P \subseteq P'$. The relation $\lb$ induces a little brother relation $\lb'$ on $\pcal'$, defined by $P' \lb' Q'$ for $P', Q' \in \pcal'$, if there exist $P, Q \in \pcal$ such that $P \subseteq P'$, $Q \subseteq Q'$, and $P \lb Q$. Let $S' \subseteq P'$ for some $P' \in \pcal'$ and put $T' = P' \setminus S'$. The set $S'$ is a splitter of $\pcal'$ with respect to $\pcal$, if for every $P \subseteq P'$ either $P \subseteq S'$ or $P \cap S' = \emptyset$, where $S' \lb' T'$ or $S'$ and $T'$ are unrelated. The splitter partition is $\pcal' \setminus \{P'\} \cup \{S', T'\}$.
By Definition~\ref{def:partialorderpartitions}, we have that $(\pcal, \lb) \lhd (\pcal', \lb')$. Note that $\pcal'$ contains a splitter if and only if $\pcal' \neq \pcal$.

Now, we can define a refinement fix-point operator $\s$. It takes as input $(\pcal_i, \lb_i) \in \pp$ and an induced parent partition pair $(\pcal_i', \lb_i')$, with $(\pcal_i, \lb_i) \lhd (\pcal'_i, \lb'_i)$, for some $i \in \bbbn$, which are stable with respect to each other. Its result are $(\pcal_{i+1}, \lb_{i+1}) \in \pp$ and parent partition  $\pcal_{i+1}'$ such that $(\pcal_{i+1}, \lb_{i+1}) \lhd (\pcal_i, \lb_i)$ and $(\pcal_{i+1}', \lb_{i+1}') \lhd (\pcal_i', \lb_i')$. Note that $\pcal_i'$ and $\pcal_{i+1}'$ differ only in one class, which is induced by the splitter that we employed to refine $\pcal_i$ to $\pcal_{i+1}$. This splitter comprises classes of $\pcal_i$, which are strict subsets from some class of $\pcal_i'$. The refinement stops, when a fix point is reached for $m \in \bbbn$ with $\pcal_m = \pcal_m'$. In the following, we omit partition pair indices, when clear from the context.

Suppose that $(\pcal, \lb) \in \pp$ has $\pcal'$ as parent with $(\pcal, \lb) \lhd (\pcal', \lb')$, where $\lb'$ is induced by $\lb$. Condition~\emph{a} of Definition~\ref{def:stabilityconditions} requires that all states in a class have or, alternatively, do not have termination options. We resolve this issue by choosing a stable initial partition pair, for $i = 0$, that fulfills this condition, i.e., for all classes $P \in \pcal_0$ it holds that either $\tss{P}$ or $\nts{P}$. For condition~\emph{b}, we specify $\lb_0$ such that $P \lb_0 Q$ with $\tss{P}$ holds, only if $\tss{Q}$ holds as well. Thus, following the initial refinement, we only need to ensure that stability conditions~\emph{c} and \emph{d} are satisfied. For convenience, we rewrite the stability conditions for $(\pcal, \lb)$ with respect to the parent partition pair $(\pcal', \lb')$. Each condition is replaced by two stability conditions, one ensuring stability of the partition and the other dealing with the little brother relation.

\begin{definition}\label{def:new-stabilityconditions}
Let $(\pcal, \lb) \in \pp$ and let $(\pcal', \lb')$ be its parent partition pair, where for all $P' \in \pcal'$ either $\nts{P'}$ or $\tss{P'}$. Then, $(\pcal, \lb)$ is stable with respect to $\pcal'$ and $B \subseteq \Sact$, if:
\begin{enumerate}

\item For all $P \in \pcal$, $a \in \Sact$, and $R' \in \pcal'$, if $P \extr{a} R'$, then $P \altr{a} \bbc{R'}{\lb'}$.

\item For all $P \in \pcal$, $b \in B$, and $R' \in \pcal'$, if $P \extr{b} R'$, then $P \altr{b} \lbc{R'}{\lb'}$.

\item For all $P, Q \in \pcal$, $a \in \Sact$, $P' \in \pcal'$, if $P \lb Q$ and $P \altr{a} R'$, then $Q \altr{a} \bbc{R'}{\lb'}$.

\item For all $P, Q \in \pcal$, $b \in B$, $R' \in \pcal'$, if $P \lb Q$ and $Q \altr{b} R'$, then $P \altr{b} \lbc{R'}{\lb'}$.

\end{enumerate}
\end{definition}
It is not difficult to observe that stability conditions~1-4 replace stability conditions~\emph{c} and \emph{d} of Definition~\ref{def:stabilityconditions}. They are equivalent when $\pcal = \pcal'$, which is the goal of our fix-point refinement operation. From now on, we refer to the stability conditions above instead of the ones in Definition~\ref{def:stabilityconditions}. The form of the stability conditions is useful as conditions~1 and~2 are used to refine the splitters and they are employed in the first phase of the algorithm, whereas conditions~3 and~4 are used to adjust the little brother relation and they are employed in the second phase. We note that if the conditions of Definition~\ref{def:new-stabilityconditions} are not fulfilled for $(\pcal, \lb) \lhd (\pcal', \lb')$, then the partition pair $(\pcal, \lb)$ is not stable.

Now, we have all the ingredients need to define the fix-point refinement operator $\s$ (for a given bisimulation action set $B \subseteq \Sact$). We define $\s(\pcal, \lb, \pcal', S') = (\pcal_r, \lb_r)$, where $(\pcal_r, \lb_r)$ is the coarsest partition pair $(\pcal_r, \lb_r) \lhd (\pcal, \lb)$ that is stable with respect to the (new) parent partition $\pcal' \setminus \{P'\} \cup \{S', T'\}$ and the stability conditions of Definition~\ref{def:new-stabilityconditions}. The existence of the coarsest partition pair $(\pcal_r, \lb_r)$ is guaranteed by Theorems~\ref{thm:lhd-lattice} and~\ref{thm:lhd-maximal}. Next, we have to show that once a stable partition pair is reached, it is no longer refined, and that $\lhd$-order is preserved by the refinement operator.

\begin{theorem}\label{thm:refinement-fixed-stable}
Let $G = (\Sst, \Sact, \tss{}, \tr{})$ and let $(\pcal, \lb) \in \pp$ over $\Sst$ be stable. For every parent partition $\pcal'$ such that $\pcal' \neq \pcal$ and every splitter $S'$ of $\pcal'$ with respect to $\pcal$, it holds that $\s(\pcal, \lb, \pcal', S') = (\pcal, \lb)$.
\end{theorem}
When refining two partition pairs $(\pcal_1, \lb_1) \lhd (\pcal_2, \lb_2)$ with respect to the same parent partition and splitter, the resulting partition pairs are also related by $\lhd$.

\begin{theorem}\label{thm:refinement-respects-lhd}
Let $(\pcal_1, \lb_1), (\pcal_2, \lb_2) \in \pp$ be such that $(\pcal_1, \lb_1) \lhd (\pcal_2, \lb_2)$. Let $\pcal'$ be a parent partition of $\pcal_2$ and let $S'$ be a splitter of $\pcal'$ with respect to $\pcal_2$. Then $\s(\pcal_1, \lb_1, \pcal', S') \lhd \s(\pcal_2, \lb_2, \pcal', S')$.
\end{theorem}
Now, taking into account Theorems~\ref{thm:lhd-lattice} -- \ref{thm:refinement-respects-lhd}, we have that iterative application of the refinement operator ultimately produces the coarsest stable partition pair.

\begin{theorem}\label{thm:refinement-good}
Let $(\pcal_c, \lb_c)$ be the coarsest stable partition pair of $G$. There exist partitions $\pcal'_i$ and splitters $S'_i$ for $i \in \{1, \ldots, n\}$ such that $\s(\pcal_i, \lb_i, \pcal'_i, S'_i)$ are well-defined with $\pcal_n = \pcal'_n$ and $(\pcal_n, \lb_n) = (\pcal_c, \lb_c)$.
\end{theorem}

We can summarize the high-level algorithm for computing the coarsest partition pair in Algorithm~\ref{alg:high-level}. The computation of the initial partition involves partitioning states to classes according to their outgoing transitions and terminations options in the vein of~\cite{FMDS2011,katoen-modelchecking}. The algorithm implements the refinement steps by splitting a parent $P' \in \pcal'$ to $S'$ and $P' \setminus S'$ and, subsequently, splits every class in $\pcal$ with respect to the splitter~$S'$ in order to satisfy the stability conditions in the vein of~\cite{fernandez,paigetarjan}. The little brother relation is adapted in the vein of~\cite{FMDS2011,report-concur}, by revisiting the little brothers of every partition class and adapting them with respect to the latest splitting. The quotient $G/(\lb \cap \lb^{-1})$ has classes $P \in \pcal$ instead of states. The termination predicate is induced by the class termination predicate as $\tss{P}$ or $\nts{P}$ for every $P \in \pcal$. The transition relation $P \tr{a} Q$ is defined according to Theorem~\ref{thm:little}. For $a \not \in B$ we have that $P \tr{a} Q$, if $P \altr{a} Q$ and there does not exist $R \neq Q$ with $Q \lb R$ such that $P \altr{a} R$. For $b \in B$ we have that $P \tr{b} Q$, if $P \altr{b} Q$ and there do not exist both $R_1, R_2 \neq Q$ with $R_1 \lb Q \lb R_2$ such that $P \altr{b} R_1$ and $P \altr{b} R_2$.

\begin{algorithm}[!t]
\caption{Computing the coarsest stable partition pair for $G = (\Sst, \Sact, \tss{}, \tr{})$ and $B \subseteq \Sact$}\label{alg:high-level}
\small
Compute the initial partition $(\pcal, \lb)$ with respect to $\pcal' = \{\Sst\}$ and $\lb' = \{(\Sst, \Sst)\}$\;

\smallskip

\While{$\pcal \neq \pcal'$}{
  Find a splitter $S'$ for $\pcal'$ w.r.t. $\pcal$\;
  $\pcal' := \pcal' \setminus \{P'\} \cup \{S', P' \setminus S'\}$\;

  Update $\lb'$\;

  Refine $\pcal$ such that it is stable w.r.t. $\pcal'$ and conditions 1 and 2 of Definition~\ref{def:new-stabilityconditions}\;

  Refine $\lb$ such that it is stable w.r.t. $\lb'$ and conditions 3 and 4 of Definition~\ref{def:new-stabilityconditions}\;

}
\smallskip

Compute the quotient $G/(\lb \cap \lb^{-1})$\;
\end{algorithm}

\newcommand{\pn}{P_{\nexists}}
\newcommand{\ppn}{P'_{\nts{}}}
\newcommand{\ppt}{P'_{\tss{}}}

\newcommand{\pcaln}{\pcal_{\nts{}}}
\newcommand{\pcalt}{\pcal_{\tss{}}}

\newcommand{\cupeq}{\mathbin{\mathord{\cup}\mathord{=}}}
\newcommand{\peq}{\mathbin{\mathord{+}\mathord{=}}}
\newcommand{\meq}{\mathbin{\mathord{-}\mathord{=}}}
\newcommand{\smeq}{\mathbin{\mathord{\setminus}\mathord{=}}}

\newcommand{\newbigbros}{\mathit{newBBNodes}}

\begin{table}
\centering
\begin{tabular}{l|rr|rrr|rrr}
\emph{Case} & \multicolumn{2}{l}{\emph{Original system}} & \multicolumn{3}{l}{\emph{Minimization by bisimulation}} & \multicolumn{3}{l}{\emph{Minimization by simulation}} \\
\emph{Name} & States & Trans & States & Trans & ExecTime(ms) &  States & Trans &ExecTime(ms) \\
\hline
CABP &464&1632&291&90&19&175&87&31\\
LIFT & 9918&4312&1299&484&235&1224&469&406\\
1BIT & 496128&81920&42723&7047&76293&9990&2628&162930
\end{tabular}
\caption{Experimental evaluation of the scalability of the algorithm for bisimulation and simulation}\label{fig:execution}
\end{table}

\section{Discussion on the Implementation and Concluding Remarks}

For the computation of the initial partition, we employ the splitting procedure in the vein of~\cite{katoen-modelchecking}, while ensuring that the little brother relation is consistent with the outgoing transitions and termination options~\cite{FMDS2011,report-concur}. We implement the first phase of the algorithm that searches for a splitter and computes the stable partition with respect to stability conditions 1 and 2 of Definition~\ref{def:new-stabilityconditions} in the vein of~\cite{paigetarjan,fernandez}. Thus, we employ the ``process the smaller half principle"~\cite{paigetarjan} as it is done for bisimulation relations~\cite{katoen-modelchecking,fernandez}. This phase of the algorithm has the same time complexity as for bisimulation, i.e., $\ocal(|\tr{}| \log |\Sst|)$~\cite{paigetarjan,fernandez,katoen-modelchecking}.

For efficient updating of the little brother relation, we give alternative representations of the sets $\lbc{P'}{\lb'}$ and $\bbc{P'}{\lb'}$ for every $P' \in \pcal'$, which are required to enforce conditions 3 and 4 of Definition~\ref{def:new-stabilityconditions}. The little brother relation~$\lb$ is kept per partition class in the form of linked lists, whereas for $\lb'$, we use a counter $\cntlb(P', Q')$ that keeps the number of pairs $(P, Q)$ for $P, Q \in \pcal$ such that $P \subseteq P'$, $Q \subseteq Q'$, $P \neq Q$, and $P \lb Q$~\cite{FMDS2011}. We keep only one Galois relation $\aetr{} = \altr{} \cup \extr{}$ and a counter $\cntal(P, a, P')$ for $P \in \pcal$, $P' \in \pcal'$ and $a \in \Sact$, where $\cntal(P, a, P')$ keeps the number of $Q'\in \pcal'$ with $P'\lb'Q'$ and $P \altr{a} Q'$~\cite{efficient-ranzato,FMDS2011}. In this way we can check the conditions of Definition~\ref{def:new-stabilityconditions} efficiently and deduce whether $P \extr{a} P'$ or $P \altr{a} P'$ whenever needed. The updating of the little brother relation has time complexity of $\ocal(|\Sact||\pcal||\lb|)$ as the little brothers are updated per label in at most $|\pcal|$ iterations of the algorithm. For space complexity we require $\ocal(|\lb|)$ for the little brother relation,  $\ocal(|\Sact||\pcal|^2\log(|\pcal|))$ is needed for the counters related to the little brother relation~\cite{efficient-ranzato,FMDS2011}, whereas in addition we require $\ocal(|\Sact||\Sst|\log(|\pcal|))$ for counters need to refine the partition~\cite{partitionpair,fernandez}, which amounts to $\ocal(|\Sst|\log(|\pcal|) + |\Sact||\pcal|^2\log(|\pcal|))$.

Finally, we implemented the algorithm~\cite{website}, and we tested it by setting the bisimulation action set to be empty and comprise all action labels and, afterwards, we compared the results with known the simulation and bisimulation minimization tools of the mCRL2 tool suite~\cite{mcrl2}, respectively.

To demonstrate the scalability of the algorithm, in Table~\ref{fig:execution}, we present experimental results for minimization with respect to bisimulation and simulation equivalences of three case studies readily available in the mCRL2 tool suite~\cite{mcrl2}: (1) the concurrent alternating bit protocol (CABP), (2) an industrial system for lifting trucks (LIFT), and (3) the onebit sliding window protocol (1BIT). It is directly observed that the minimization by bisimulation is much more time-effective. The prototype implementation, however, relies on linked lists, which overhead reflects into the execution time for minimization by bisimulation.

%


\bibliographystyle{eptcs}
\bibliography{references}
\end{document}